# Converging trend of global urban land expansion sheds new light on sustainable development

Shengjie Hu[1,2,3], Zhenlei Yang[2,3], Sergio Andres Galindo Torres[2,3], Zipeng Wang[4], Haoying Han[5], Yoshihide Wada[6,7], Thomas Cherico Wanger[2,3,8]*, Ling Li[2,3]*

**Affiliations:**

[1]College of Environmental and Resource Sciences, Zhejiang University; Hangzhou, 310058, China.

[2]School of Engineering, Westlake University; Hangzhou, 310024, China.

[3]Key Laboratory of Coastal Environment and Resources of Zhejiang Province, Westlake University; Hangzhou, 310024, China.

[4]School of Science, Westlake University; Hangzhou, 310024, China.

[5]College of Civil Engineering and Architecture, Zhejiang University; Hangzhou, 310058, China.

[6]Climate and Livability, Center for Desert Agriculture, Biological and Environmental Science and Engineering Division, King Abdullah University of Science and Technology; Thuwal, 23955-6900, Saudi Arabia.

[7]Biodiversity and Natural Resources Program, International Institute for Applied Systems Analysis (IIASA); Laxenburg, A-2361, Austria.

[8]Agroecology, Department of Crop Sciences, University of Göttingen; Grisebachstr. 6, D-37077 Göttingen, Germany.

*Corresponding author. Email: liling@westlake.edu.cn; tomcwanger@westlake.edu.cn

**Abstract:** Urban land growth presents a major sustainability challenge, yet its growth patterns and dynamics remain unclear. We quantified urban land evolution by analyzing its statistical distribution in 14 regions and countries over 29 years. The results show a converging temporal trend in urban land expansion from sub-country to global scales, characterized by a coherent shift of urban area distributions from initial power law to exponential distributions, with the consequences of reduced system stability and resilience, and increased exposure of urban populations to extreme heat and air pollution. These changes are attributed to the increased influence from external economies of scale associated with globalization and are predicted to intensify in the future. The findings will advance urban science and direct current land urbanization practices toward sustainable development, especially in developing regions and medium-size cities.

One of the most defining changes that humanity has brought about since the Industrial Revolution is the rapid replacement of natural land with urban structures to support the growing number of urban dwellers and their demands[1, 2]. Urban growth indeed facilitated socio-economic development and improved human living conditions, sometimes summarized as urban economies of scale[3,4]. However, it also caused problems that directly threaten human health, such as the urban heat island effect (UHI) and fine particulate matter (PM2.5) pollution[5,6]. UHI, coupled with global warming, has increased the exposure of global urban populations to extreme heat by 200% from 1983 to 2016[5]. Nearly 86% of global urban inhabitants lives with unhealthy PM2.5 levels and the threat of PM2.5 is increasing globally as urban land continues to grow[6]. For current environmental and sustainability issues, how human societies use and manage land is both a source of problems and solutions[7]; therefore, the future trajectory of land urbanization will be closely linked with the sustainable development of humanity[8].

Effectively addressing the threats and risks associated with land urbanization lies in understanding the growth dynamics of urban land and its impacts on urban systems and their residents in order to better inform urban planning and management[1,3,8]. Since most characteristics of cities are contingent on their size, studying the size distribution of ensemble of cities (i.e., the urban system) and its spatio-temporal variations has been considered to provide insights into the dynamics of urban systems[3,4,9]. However, the current statistical lens of urban science is mainly concentrated on urban population[3,10,11], leaving almost a gap in urban land area, which is also an important proxy of city size, except for the amount and rate of area change[12]. Moreover, cities emerge from the aggregation of people, generically driven by economies of scale that consist of internal and external parts[3,4,13]. Globalization, another defining feature of our times, has strengthened the connectedness, interdependence and integration among cities, resulting in an increased influence from external economies of scale at the city level[14,15]. But how urban systems in regions with different political, economic and geographical conditions respond to this driving force change remains unknown, especially in the land dimension. Furthermore, despite urban land growth is known to raise the burden of disease, the specific relationship between its changing patterns and configurations, and the health risks (e.g., UHI and PM2.5 pollution) it poses still requires more research to be determined[16]. Therefore, an improved understanding of urban land expansion is necessary and urgent to create a healthy and sustainable living space for humanity.

Here, we aim to address the above key knowledge gaps in urban science by quantifying the evolutionary process of urban land expansion (Fig. 1). We analyzed the urban land area distribution and its changes from 1992 to 2020. With comparison of the results in 14 regions covering global, continental, national and sub-country scales, we determined the universal distribution function and trends of changes. Using fluctuation, entropy and power spectrum analyses, we connected the temporal variations of urban area distributions to the alterations of system state and functions. We explained the evolutionary characteristics in urban systems through the changing roles of internal and external economies of scale. We discussed the implications of our findings on how to stir current land urbanization practices for sustainable development by linking urban system changes to altered exposure of urban inhabitants to extreme heat and PM2.5 pollution, and by predicting the future trajectory of urban land growth.

## Results and Discussion
### Convergent evolution of urban land

Statistical analysis of urban land areas was performed for cities within 14 regions, including the globe, each of the continents except Antarctica, the United States of America (USA), China, India, and four economic zones of China (Supplementary Fig. 1). These study regions were selected to cover various spatial scales as well as different urbanization modes and stages, and levels of economic development. The cities discussed here are not defined based on administrative boundaries but spatial clusters with dense human activities, because the latter could more realistically reflect the local socio-economic conditions[17,18]. The results showed that the urban areas all follow well the shifted power law distribution (equations (1)(3b)) over the whole study period of 29 years (1992-2020) (Fig. 2 and Supplementary Figs. 2-15). The three coefficients involved in the distribution function, increasing over time, are interrelated within each region; and such interrelations can be described by the same functions in all regions (Fig. 2, Fig. 3A and B, and Supplementary Figs. 16-18). Specifically, the shift coefficient ($b$), taken as an independent parameter, increases with time exponentially (equation (3c), Fig. 2 and Supplementary Fig. 16), whereas the scaling exponent ($a$) and the proportion term ($c$) are linearly and quadratically related with $b$, respectively (equations (3d)(3e), Fig. 2 and Supplementary Fig. 17-18). Thus, the temporal evolution of urban area distributions can be recapitulated in a unified equation, $P(A,t) = c(b)[A + b(t)]^{-a(b)}$ with $b(t) \propto e^{c_1 t}$. The exponential increase of $b$ indicates that the urban area distribution evolves along the direction away from its initial power law distribution ($b = 0$) at an accelerated rate and will approach an exponential and eventually uniform distribution ($b \to +\infty$) in the future (equation (2) and Supplementary Text1).

The changes in urban land arrangement corresponding to the evolution of urban area distribution during the study period were explored by analyzing the alterations in city configuration in terms of the proportional composition of different types of cities in a region. Four city types were defined according to UN-habitat[19], including small-size cities ($A < 10^1 \text{km}^2$), medium-size cities ($10^1 \leq A < 10^2 \text{km}^2$), large-size cities ($10^2 \leq A < 10^3 \text{km}^2$) and mega-size cities ($A \geq 10^3 \text{km}^2$) (Supplementary Text3). We found that the small-size cities dominated all regions at the beginning (1992), with higher proportions in developing regions like China (92%) than that in developed regions like USA (64%) (Table 1 and Supplementary Table 1). The medium-size cities then gradually took the lead and accounted for around 60% worldwide by 2020. The portions of large- and mega-size cities also increased and showed their highest values in North America (NA). These results demonstrated that the city configuration changes in the same manner and gradually becomes similar among regions, as evidenced by the collective shift of the dominant city type from small- to medium-size cities and reduced regional disparities in the proportion of each type. The increase in the $b$ value of the urban area distribution corresponds to the increase in the proportion of medium-size cities in the city configuration.

As shown above, urban land areas in different regions follow the same distribution function, evolve along the same direction, undergo the same pattern shift in city configuration and exhibit a converging trend of changes across spatial scales and over time, irrespective of political, economic, cultural or geographical conditions (Fig.2, Supplementary Figs. 19-20, and movie S1). The convergent evolution of urban land could be attributed to three factors: (1) Human beings, as one species, differ in details but are alike in essentials[20], including similar demands in habitat development and organization; (2) Urbanization level is assessed by the same standard across regions, that is, an urbanized region has an urban population over 70%[21]. This may lead to the same destination of land occupation; and (3) Urbanizing countries tend to learn from urbanized countries, which seems to be supported by the phenomenon that the area distributions all moved towards USA's situation (Fig. 2 and Supplementary Fig. 20). It seems to be a surprising result that city configurations all became dominated by the medium-size cities and gradually approached to the same proportional composition at the national scale (Table 1 and Supplementary Table 1), considering the very different policies implemented by countries. For example, USA encourages the free-market economy and imposes no restrictions on city size[22], whereas China intentionally restricts the size of its large cities[23]. One possible reason is that the rapid urbanization process in developing countries promoted the establishment of small-size cities and their growth to the medium-size ones[23], while the counter-urbanization and re-urbanization processes undergone by developed countries have a similar effect on driving the dominance of medium-size cities in the city configuration[22]. A more fundamental reason probably lies in the dependence of economies of scale on city size, i.e., economies of scale may not work if the city is too small, whilst diseconomies of scale may arise if the city is too large[24].

**Changes in the state and functions of urban systems**

(1) State change and phase transition

To explore how the state of urban systems has changed over time as reflected by the shift of urban area distributions, we conducted fluctuation and entropy analyses, which are widely used to quantify the system state in statistical physics. Two indicators, coefficient of variation (CV, equation (4)) and information entropy (H, equation (5)) that respectively measure the degrees of heterogeneity and chaos of a system[25,26], were calculated for each region. A trend with decreasing CV and increasing H emerges from all regions (Figs. 3C and D), demonstrating that as the urban area distributions deviated from their initial power law distributions, the urban systems developed towards a relatively homogeneous and disordered state in the land dimension.

By further analyzing the distribution function of urban areas, we found a phase transition that is responsible for the change in system state (Supplementary Fig. 19). Specifically, the functional properties of the shifted power law distribution suggest that urban area distributions can be decomposed into two components according to the values of $b$ (equation (6)). Statistical fitting subsequently showed that the one composed of cities with area smaller than $b$ obeys the stretched-exponential distribution (equation (7), Supplementary Figs. 21-34), whereas the other one follows the power law distribution (equation (8), Supplementary Figs. 35-48). Analogous to the concepts of state and phase in statistical physics[27], we refer to these two distributions as the two phases of urban system state. The balance and competition of the two phases in the area distribution were quantified by calculating the proportions of cities locating in each component, denoted as $R_1$ (power law phase, equation (9a)) and $R_2$ (stretched-exponential phase, equation (9b)). A trend of expanding stretched-exponential phase and meanwhile contracting power law phase over time was observed in all study regions (Fig. 3E and Supplementary Fig. 19). As shown by previous studies, the stretched-exponential distribution represents a homogenous, disordered state[28], while the power law distribution indicates a heterogeneous, ordered state[29,30]. Combining these results, we suggested that a phase transition, characterized by the expansion of disordered phase and the contraction of ordered phase, took place during the evolution of urban land areas, driving the urban systems towards a homogeneous and disordered state. The increase of $b$ value signals the state change of urban system and indicates the degree of such change.

(2) Reduced system stability and resilience

The decreasing trend of CV (Fig. 3C) found in urban land areas indicates that the size diversity (i.e., number of distinct urban areas) of urban systems has declined. Previous studies showed that the size of a city affects how it responds to disturbances[31] and the decline of size diversity may result in a loss of response diversity that can destabilize the system[32]. Moreover, medium-size cities were found to be more vulnerable to extreme events than mega cities[31]. To quantitatively measure the stability changes in urban systems, the effective potential (EP, Eq. 10) was introduced, because the most stable state of a system is the one with the minimum EP[33]. Since the value of EP is positively correlated with $b$ (equation (10)), the increasing trend of $b$ values demonstrates that the stability of urban systems is weakening. Thus, the decline in size diversity and the dominance of medium-size cities does reflect the reduced stability of urban systems.

To further quantify the functional alterations of urban systems, power spectrum analysis[34] was performed to the ordered power law phases by calculating urban area power spectrum (equation (11)), the corresponding autocorrelation function[35] (equations (12)(13)) and the Hurst exponent[36] ($H_{urst}$, equation (14)). The values of spectrum exponent ($\gamma$) were found to increase throughout the study period and change from negative values in early years to positive ones for later/recent years (Fig. 3F). In signal analysis, negative and positive values of $\gamma$ are associated with pink and blue noise[37] (Supplementary Text5), respectively. Hence, all regions had "pink" power spectrums for the power law distributed urban areas in the beginning, which then turned "blue" and became "bluer" over time (Fig. 3F). With this "blue shift" tendency of area power spectrum, the autocorrelation function (i.e., the Fourier counterpart of power spectrum) exhibits a possibility of approaching infinity (equation (13d)). Since the increased autocorrelation is an early-warning signal of critical tipping[38], the "blue shift" trend of urban areas may also indicate a loss of system resilience, meaning that it will take urban systems longer time to recover from disturbances[38]. Moreover, the values of $H_{urst}$ also increased and exceeded 0.5 in most regions since 2014 (Fig. 3G), suggesting that the driving processes behind these trends have begun to operate in a persistent mode[39]; thus, the possible alterations of system functionality may continue or even become more severe in the future.

**Mechanism underlying changes of urban systems**

Economies of scale are a widely accepted driver of population agglomeration and city formation[3,4,13]. Here, it refers to the benefits obtained by a city when the urban area increases, and its internal and external parts are the benefits brought by the urban land growth inside and outside the city, respectively. It is difficult to isolate and quantify these two parts in practice, because they are intertwined and affected by numerous factors. However, since the connectedness and interdependence among regions has been strengthened by globalization[14,15,21], the globalization index[40] (GI) can be used as an alternative indicator to represent the degree to which a region is influenced by external economies of scale. The rising GI values (Fig. 3H) reflect that the study regions have been more closely connected over the past 30 years; hence, the influence from external economies of scale on each region has been intensified.

The shift coefficient ($b$), which represents the evolutionary characteristics of urban systems, correlates positively with GI values in all regions (Fig. 3I), suggesting that the increased influence from external economies of scale is one of the driving factors that give rise to the variations of the urban area distribution as well as the alterations of system state and functions over the study period. Thus, the phase transition in urban area distributions characterized by the contraction of power law phase and expansion of stretched-exponential phase can be linked to the shift of driving force from internal to external. This correspondence is particularly evident in China, where the urban area distribution contained solely power law phase in 1992 but the stretched-exponential phase overtook it and accounted for over 60% by 2020 (Figs. 2 and 3E). Although China opened up to the outside since 1978, the degree of openness was low in (and prior to) 1992 (Fig. 3H) and the interactions among the cities within the country were also weak due to the underdeveloped transportation infrastructure and restrictions imposed by the "hukou" policy (urban residence permit)[21]. Therefore, internal economies of scale (internal force) played a dominant role in 1992 and the urban areas followed the power law distribution. After joining WTO in 2001, the influence from external economies of scale (external force) gradually intensified and thus the power law phase shrank, while the stretched-exponential phase appeared and expanded under the mixed effect of internal and external economies of scale. Since urban areas will approach an exponential distribution and eventually a uniform distribution as $b$ goes to positive infinity, we speculated that an exponential area distribution will emerge when external economies of scale play a dominant role and a uniform area distribution if it completely dominates (Fig. 1).

The power law phases of different regions were found to collapse together in each year, when the urban areas are rescaled with $b$ (Fig. 1 and Supplementary Fig. 50). As $b$ is the only independent parameter of the distribution function and is time-dependent, these results are reminiscent of the dynamical scaling phenomenology shared by coarsening systems – the domain morphology remains statistically the same at all times when rescaled by a single characteristic scale that grows temporally[41]. Therefore, coarsening dynamics that system relaxes along the steepest gradient in its energy landscape[42] may be the dynamical principle governing the expansion of urban land.

**Implications for sustainable development**

(1) More severe and persistent in developing regions

Despite the consistency in overall trends, regional distinctions still exist, especially between developed and developing regions, as shown by the different values of distribution function coefficients and system state indicators across regions and over the years (Fig. 3). In general, the values of these quantities are larger in developed regions, especially before 2014 (Fig. 3), resulted from the fact that developed regions completed urbanization before 1992 while developing regions are still in the process of urbanization. However, the magnitude of value change between 1992 and 2020 is more dramatic in developing regions (Table 2), indicating that the degree to which the urban system state (CV and H) and functions ($\gamma$ and $H_{urst}$) have changed differs among the study regions and tends to be greater and more persistent in developing regions (See Supplementary Text6 for a detailed discussion).

(2) Importance of medium-size cities in mitigating UHI and PM2.5 pollution

Avoidance of the urban heat island effect (UHI) and air pollution with downstream health implications for urban inhabitants is of major concern to the UN Sustainable Development Goal of "reduce adverse effects of natural disasters"[43]. The intensity of UHI has been found to be positively correlated with urban area and the relationship could be described by a sigmoid function[44] (Fig. 4A), in which we observed the fastest growth in UHI intensity in cities with area between $10^1$ and $10^2$ km$^2$ (i.e., the medium-size cities defined here). As the dominant city type in urban systems has shifted from small- to medium-size cities, the current path of land urbanization tends to exacerbate UHI. By incorporating UHI data[5] into our analysis, we found a positive correlation between the total urban population exposure and the shift coefficient ($b$) for all study regions (Fig. 4A), indicating that as the urban systems evolve, the risk of their dwellers being exposed to extreme heat has increased. Except for harming public health, urban heat stress also caused labor losses[45]. Since the low-paid sector was more vulnerable to this consequence[45], the changes in urban systems may also indirectly contribute to income inequality. However, as the other side of the coin, these results also point out that effective control of UHI in the medium-size cities would help mitigate these negative effects.

When considering PM2.5 pollution[46], we found that the highest PM2.5 concentrations globally and in developing regions occur in the medium-size cities, while in developed regions they occur in the small-size ones (Fig. 4B). Therefore, the shift of dominant city type from small- to medium-size cities could exacerbate the PM2.5 pollution in developing regions but mitigate it in developed regions. This result not only partly explains the opposite trends observed from the raw PM2.5 data[46], but also implies that effective treatment of PM2.5 pollution in the medium-size cities would improve the air quality in all regions.

(3) Future trajectory of land urbanization

The shifted power law distribution with the three temporally varying coefficients provides a model for predicting the future trajectory of urban land growth. The statistical distributions of urban areas across the world will collectively move further away from the initial power law distributions, a tendency that will be even more pronounced in developing regions (Table 3). The portion of the ordered power law phase in the urban area distribution will only account for 0.11% (India) to 20.75% (NA) in 2050 and completely disappear in most regions by 2100 (Table 3). After that, the urban areas in all regions will probably enter the regime of exponential distribution (equation (2b) and Fig. 1), which represents a random state[47]. However, given that all regions have experienced an abrupt increase of $b$ value during 2013-2014 (Fig. 3A), the urban systems globally may reach the predicted situation earlier. Since the shifted power law distribution will be approximate to a uniform distribution as $b$ becomes sufficiently large (equation (2c) and Fig. 1), the urban systems could eventually approach a thermodynamic equilibrium-like state with cities of all sizes appearing with equal probability[48] if the current urbanization scenario continues. Such an evolutionary trend indicated by the statistical distribution shifts may correspond to a further decrease in urban system stability and resilience, as well as an increase in the risk of exposure to hazards. Therefore, it is necessary to rethink the current practices of urbanization, especially the share of medium-size cities in urban planning, to achieve sustainable development. Given that urban evolution is closely linked to social, economic and political factors that influence the land development and organization practices[1,16], future research should focus on unifying these dimensions of urbanization to advance urban science with predictive theories and models (Supplementary Text7).

## Methods

### Study regions and data sources

This study was carried out in 14 countries and regions, including the globe, continents other than Antarctica, USA, China, India and the four economic zones of China (Supplementary Fig. 1). These regions were chosen to cover a) various spatial scales, from local to global, considering that the urbanization process unfolds on multiple scales[1]; b) diverse political, economic, social and cultural backgrounds, as the practice of land urbanization is influenced by these aspects[1,16]; and c) different urbanization stages[22], exemplified by the counter-urbanization and re-urbanization process undergone by developed regions like Europe and USA, and the rapid urbanization process experienced by developing regions such as Africa and China; thereby revealing common features and patterns of urban systems with general implications for urban researchers and policy-makers in different regions.

Urban land areas in each region were extracted from a global dataset of annual urban extents from 1992 to 2020 with 1km resolution[17]. The dataset was developed based on the harmonized time-series nighttime light data facilitated by data of water masks, global artificial impervious area and etc. in a stepwise-partitioning framework integrating multiple algorithms[17]. It has been cross-evaluated with other global urban products, historical google maps and socioeconomic statistics, and shown to be reliable[17]. Nonetheless, nighttime light data may be affected by national or local energy efficiency or other control policies. Such effect is difficult to quantify, but relatively small and can be neglected in our long-term analysis. Since the dataset delineated city boundaries by spatial clustering instead of administrative divisions, the urban area data it offered could more realistically reflect the locally high-intensity human activities[17,18]. The long-term urban land dataset was established for each study region with the assistance of ArcGIS 10.2.

Globalization index (GI), measuring the globalization degree of a region, was used to indirectly reflect the influence of external economies of scale in this study. The KOF Globalization Index was used because it measures the economic, social and political dimensions of globalization and has a long temporal coverage from 1970 to 2019[40]. Since this is a country-level dataset, the GI data for each continent was calculated as the average of all countries on that continent.

The impacts of urban system alterations on urban dwellers were analyzed in terms of urban heat island effect (UHI) and particulate matter (PM2.5) pollution based on datasets of Global High Resolution Daily Extreme Urban Heat Exposure (UHE-Daily, 1983-2016)[5] and The Annual PM2.5 Concentrations for Countries and Urban Areas (1998-2016)[46], respectively. These two hazards were analyzed because they are closely associated with urban land expansion and are among the leading contributors to global burden of disease[5,6], and their data coverage is consistent with the urban extent dataset. By spatially overlaying the UHE-Daily with the boundaries of the 14 study regions, the total annual exposure of urban population within each region was determined. Based on the city configuration we defined and the PM2.5 data, the average PM2.5 concentrations for each city type were calculated using area as a weight. Because of the limited sample size of these two datasets, these analyses were conducted at global, continental and national scales.

**Urban area distribution**

The shifted power law distribution, $f(x)$ (equation (1a)) was proposed by Mandelbrot by adding a shift coefficient $b$ ($b \geq 0$) to the power law distribution[49]. The definition of it is as follows,

$$f(x) = C(x+b)^{-\alpha} \qquad (1a)$$

With

$$\alpha > 1, b \geq 0 \; and \; C = C(\alpha, b, x_{min}, x_{max}) > 0 \qquad (1b)$$

Satisfying,

$$\int_{x_{min}}^{x_{max}} f(x)dx = \int_{x_{min}}^{x_{max}} C(x+b)^{-\alpha}dx = 1 \qquad (1c)$$

where $x$ is the variable of interest; $x_{max}$ and $x_{min}$ are the upper and lower limits of the sample set of $x$; $f(x)$ denotes the probability density function of $x$; $\alpha$, $b$ and $C$ are the coefficients of $f(x)$ called the scaling exponent, shift coefficient and proportion term, respectively. The specific functional form of $C$ is given in Supplementary Text1 equation (S2a).

The shifted power law distribution exhibits different behaviors as the value of $b$ increases, i.e., it reduces to a power function at $b = 0$ (equation (2a)) and can be approximated by the exponential (equation (2b)) and uniform distribution (equation (2c)) as $b$ approaches positive infinity. A detailed derivation is given in Supplementary Text1.

$$if\ b = 0, f(x) \propto x^{-\alpha} \tag{2a}$$

$$if\ b \to +\infty, \alpha \to +\infty, and\ \frac{\alpha}{b} = r, f(x) \propto e^{-rx} (r > 0) \tag{2b}$$

$$if\ b \to +\infty, f(x) \approx \frac{1}{x_{max}-x_{min}} (x_{max} > x_{min} > 0) \tag{2c}$$

where $r$ is a constant number.

The shifted power law distribution was fitted to the complementary cumulative density function (CCDF, equation (3b)) of urban land areas (i.e., the urban area distribution) in each study region. CCDF was used rather than the probability density function (PDF, equation (3a)) because it is more robust against the fluctuation caused by finite system size[50]. The coefficients of urban area distribution vary among regions and increase with time. The temporal variation of the shift coefficient ($b$) can be fitted by exponential functions in all regions (equation (3c)), while the scaling exponent ($a$) and the proportion term ($c$) are related with $b$ by linear and quadratic functions (equations (3d) (3e)), respectively.

$$p(A) = \Pr(X = A) = C(A + b)^{-\alpha}\ (\alpha > 1;\ b \geq 0;\ C > 0) \tag{3a}$$
$$P(A) = \Pr(X \geq A) = c(A + b)^{-a}\ (a = \alpha - 1 > 0;\ c > 0) \tag{3b}$$
$$b(t) = c_1 e^{c_1' t}(t = Year - 1992) \tag{3c}$$
$$a(b) = c_2 + c_2' b \tag{3d}$$
$$c(b) = c_3 + c_3' b + c_3'' b^2 \tag{3e}$$

where $A$ represents the urban land area ($A \in [A_{min}, A_{max}]$ with the unit of km$^2$), which is the statistical variable we are interested in; $X$ is the observed value of $A$; $p(A)$ and $P(A)$ are the PDF and CCDF of $A$, respectively, and the latter is the urban area distribution we are studying; $\alpha$ and $a$ are the scaling exponents of the PDF and CCDF for $A$, respectively, and $a$ ranges from 0 to 2 according to the fitting results; $b$ denotes the shift coefficient; $t$ is the time index, starting from zero (i.e., year of 1992); $C$ and $c$ represent the proportion terms which are constrained by the full probability condition; $c_1$, $c_1'$, $c_2$, $c_2'$, $c_3$, $c_3'$ and $c_3''$ are fitting parameters.

Based on the exponential increase of the shift coefficient ($b$), it can be inferred that the urban area distribution ($P(A)$, equation (3b)) will approach the exponential and uniform distribution in the future (equations (2b) (2c)). Since the scaling exponent ($a$) and the shift coefficient ($b$) of urban area distributions are linearly correlated (equation (3d)), and the total area of urban land accounts for less than 7% of the regional area in each study region (Supplementary Table 2), the urban area distribution in all study regions is likely to enter the regime of the exponential distribution first in the near future (the year of 2100, Table 3).

All fittings in this work were performed through the scipy.optimize.curve_fit model in python 3 (https://www.python.org/) and the fitting performance was assessed by the Coefficient of Determination (r$^2$) and Root Mean Squared Error (RMSE). See Supplementary Text2 for details.

**Quantification of urban system state**

The coefficient of variation (CV, equation (4)), defined as the ratio of the standard deviation ($\sigma$) to the mean ($\mu$), can reflect the degree of heterogeneity of a system, i.e., the larger the CV value, the higher the degree of heterogeneity[25]. Information entropy (H, equation (5)) was calculated to determine the degree of chaos of the system state and a higher value of H indicates a more chaotic system state[26].

$$CV = \frac{\sigma}{\mu} \tag{4}$$

$$H = -\sum_{i=1} p_i(A) \log_2 p_i(A) \tag{5}$$

where $\sigma$ and $\mu$ are the standard deviation and mean of the urban area ($A$) of an urban land dataset, respectively; and $p_i(A)$ represents the probability of the $i$-th urban area in the whole dataset.

**Phase determination**

Two components, a flat "shoulder" at the lower limit and a power law-like "body" at the upper limit, were observed to coexist in the urban area distributions shown on the log-log plot (Supplementary Fig. 19). To separate the two components for further analysis of their characteristics, we compared the shifted power law distribution and the power law distribution that has the same coefficients as it (equation (6)). We found that the difference of urban areas on the two distributions mainly depends on their relationships with the value of $b$ as shown by equation (6c). When the urban areas are larger, the effect of $b$ can be neglected and the distribution approximately follows a power law; otherwise, the influence of $b$ is evident, causing urban areas smaller than $b$ to deviate from the power law. Thus, the $b$ value can be taken as the critical area that separates the two components of the urban area distribution.

$$y_1(A) = \log[c(x+b)^{-a}] = c - a\log(A+b) \tag{6a}$$

$$y_2(A) = \log(cx^{-a}) = c - a\log(A) \tag{6b}$$

$$\Delta y(A) = |y_1(A) - y_2(A)| = |a\log(\frac{1}{1+b/A})| \tag{6c}$$

where $y_1(A)$ and $y_2(A)$ are log-transformed CCDF functions corresponding to the shifted power law and the power law distribution of the urban area ($A$), respectively, and $\Delta y(A)$ denotes the difference between them.

The year-by-year urban land data in each region was then partitioned into two subsets: the shoulder subset consists of the cities with areas smaller than $b$ and the body subset contains all remaining cities, with the sample size denoted as $N_1$ and $N_2$, respectively. The shoulder subset was found to fit well the stretch-exponential distribution (equation (7)), while the body subset conforms to the power law distribution (equation (8)), and their respective proportions in an urban area distribution given by $R_2$ (stretched-exponential, equation (9b)) and $R_1$ (power law, equation (9a)),

$$P_{se}(A) = \Pr(X \geq A) = C_1 A^{\epsilon-1} e^{-C_1' A^\epsilon} (A_{min} \leq A < b; \epsilon < 0) \tag{7}$$

$$P_{pl}(A) = \Pr(X \geq A) = C_2 A^{-a_{pl}} (b \leq A \leq A_{max}; a_{pl} > 0) \tag{8}$$

$$R_1 = \frac{N_2}{N_1 + N_2} \tag{9a}$$

$$R_2 = \frac{N_1}{N_1 + N_2} \tag{9b}$$

where $P_{se}(A)$ and $P_{pl}(A)$ denote the stretch-exponential and the power law distribution of the urban area ($A$), respectively; $X$ is the observed value of $A$; $\epsilon$ represents the exponent of the stretched-exponential distribution; $a_{pl}$ is the power exponent of the power law distribution with values between 0 and 2, depending on the results of the fit; $C_1, C_1'$, and $C_2$ are fitting parameters.

**Effective potential**

Effective potential (EP) is a measure of the macro-state stability of a system by means of its probability density function (PDF) and the most likely state of the system is the one with the minimum value of EP[33]. According to the definition of EP (equation (10)), the PDF of urban area ($p(A)$, equation (3a)) was used to derive the EP of the urban system. The result, equation (10), showed that the EP value of the urban system is positively related to the shift coefficient ($b$), indicating that the stability of the urban system will decrease as the value of $b$ increases.

$$EP = -\log(p(A)) = \alpha \log(A+b) - \log(C) \tag{10}$$

where $p(A)$ is the PDF of the urban area ($A$); $\alpha$, $b$ and $C$ are the function coefficients, the same as in equation (3a).

**Power spectrum analysis**

(1) Area power spectrum

The area power spectrum, $S(\varphi)$, corresponding to the power law phase (equation (8)) of the urban area distribution was established through the transformation used in Bak et al. (ref. 34), given below,

$$S(\varphi) = \int_b^{A_{max}} \frac{A p_{pl}(A)}{1+(\varphi A)^2} dA \propto \varphi^\gamma \quad (\gamma = a_{pl} - 1) \tag{11}$$

where $A$ is the urban area of the power law phase ($A \in [b, A_{max}]$); $\varphi$ represents the transformed quantity corresponding to $A$; $p_{pl}(A)$ denotes the PDF corresponding to the power law phase of the urban area distribution; $a_{pl}$ is the power exponent of the power law phase; and $\gamma$ is the exponent of the power spectrum. See Supplementary Text4 for a detailed derivation.

The value of spectrum exponent $\gamma$ indicates the type of power spectrum[37]. If $\gamma = 0$, the power spectrum is a white noise type; if $\gamma < 0$, it is classified as the pink noise; if $\gamma > 0$, it belongs to the blue noise. More information about the definitions and properties of these "colored" noises can be found in Supplementary Text5.

(2) Autocorrelation function

Autocorrelation function ($R(\tau)$, equation (12a)), as the Fourier counterpart of power spectrum, can be obtained by performing the Fourier transform to the power spectrum ($S(\varphi)$)[35]. When $R(\tau)$ and $S(\varphi)$ are even functions, their relationship can be simplified as equation (12b).

$$R(\tau) = \frac{1}{2\pi} \int_{-\infty}^{+\infty} S(\varphi) e^{i\varphi\tau} d\varphi \tag{12a}$$

$$R(\tau) = \frac{1}{\pi} \int_0^{+\infty} S(\varphi) \cos \varphi\tau \, d\varphi \tag{12b}$$

Based on the assumption of even functions, the analytical expression of $R(\tau)$ (equation (13a)) was derived and its behaviors under different types of $S(\varphi)$ were analyzed. The results demonstrated that when $S(\varphi)$ is a white noise ($\gamma = 0$), $R(\tau)$ is a $\delta$ function (equation (13b)); when $S(\varphi)$ belongs to pink noise ($\gamma < 0$), $R(\tau)$ goes asymptotically to zero (equation (13c)); and when $S(\varphi)$ belongs to blue noise ($\gamma > 0$), $R(\tau)$ diverges, with a possibility of approaching infinity (equation (13d)). According to the critical tipping theory, increasing $R(\tau)$ indicates the loss of system resilience and potential transition of system state[38]. Therefore, the "blue shift" of urban area power spectrums, indicated by $\gamma$ turning from negative to positive values, is likely to signify a decline of system resilience.

$$R_A(\tau) = \frac{1}{\pi} \int_0^{+\infty} \varphi^\gamma \cos \varphi\tau \, d\varphi$$

$$= \frac{1}{\pi\tau}\varphi^\gamma \sin(\varphi\tau)|_{\varphi\to+\infty} + \frac{\gamma}{\pi\tau^2}\varphi^{\gamma-1} \cos(\varphi\tau)|_{\varphi\to+\infty} + \frac{\gamma(\gamma-1)}{\pi\tau^3}\varphi^{\gamma-2} \sin(\varphi\tau)|_{\varphi\to+\infty}$$

$$+ \frac{\gamma(\gamma-1)(\gamma-2)}{\pi\tau^3}\int_0^{+\infty} \sin(\varphi\tau)\varphi^{\gamma-3}d\varphi \quad (-1 < \gamma < 1) \tag{13a}$$

$$if\ \gamma = 0,\ R(\tau) = \frac{1}{\pi\tau}\sin(\varphi\tau)|_{\varphi\to+\infty} = \delta(\tau) \tag{13b}$$

$$if\ \gamma < 0,\ R(\tau)|_{\varphi\to+\infty} \to 0 \tag{13c}$$

$$if\ \gamma > 0,\ R(\tau) \approx \frac{1}{\pi\tau}\varphi^\gamma \sin(\varphi\tau)|_{\varphi\to+\infty}\ (\text{diverge}) \tag{13d}$$

(3) Hurst exponent

Since the urban area power spectrum belongs to the fractional Gaussian noise ($-1 < \gamma < 1$, Fig. 3F), the Hurst exponent ($H_{urst}$) can be calculated using equation (14)[36]. According to the fractional Brownian motion theory[39], the value of $H_{urst}$ could reflect the coupling mode of the processes that drive the changes of urban area growth and statistical distribution. To be specific, $H_{urst} = 0.5$ indicates no correlation between processes, $H_{urst} > 0.5$ indicates that the correlation between processes is persistent and $H_{urst} < 0.5$ indicates that the correlation is anti-persistent[39].

$$H_{urst} = \frac{\gamma-1}{2} + 1 \tag{14}$$

**Data availability**

All data used are publicly available. The global annual urban extents dataset was downloaded from https://doi.org/10.6084/m9.figshare.16602224.v1. Globalization index (GI) data is provided by KOF Swiss Economic Institute and can be accessed through https://kof.ethz.ch/en/forecasts-and-indicators/indicators/kof-globalisation-index.html. Datasets of Global High Resolution Daily Extreme Urban Heat Exposure (UHE-Daily, 1983-2016) and Annual PM2.5 Concentrations for Countries and Urban Areas (1998-2016) are distributed by the Socioeconomic Data and Application Center and can be downloaded from https://sedac.ciesin.columbia.edu/data/set/sdei-high-res-daily-uhe-1983-2016 and https://sedac.ciesin.columbia.edu/data/set/sdei-annual-pm2-5-concentrations-countries-urban-areas-v1-1998-2016, respectively.

**Acknowledgments**

This work was supported by National Natural Science Foundation of China 41976162.

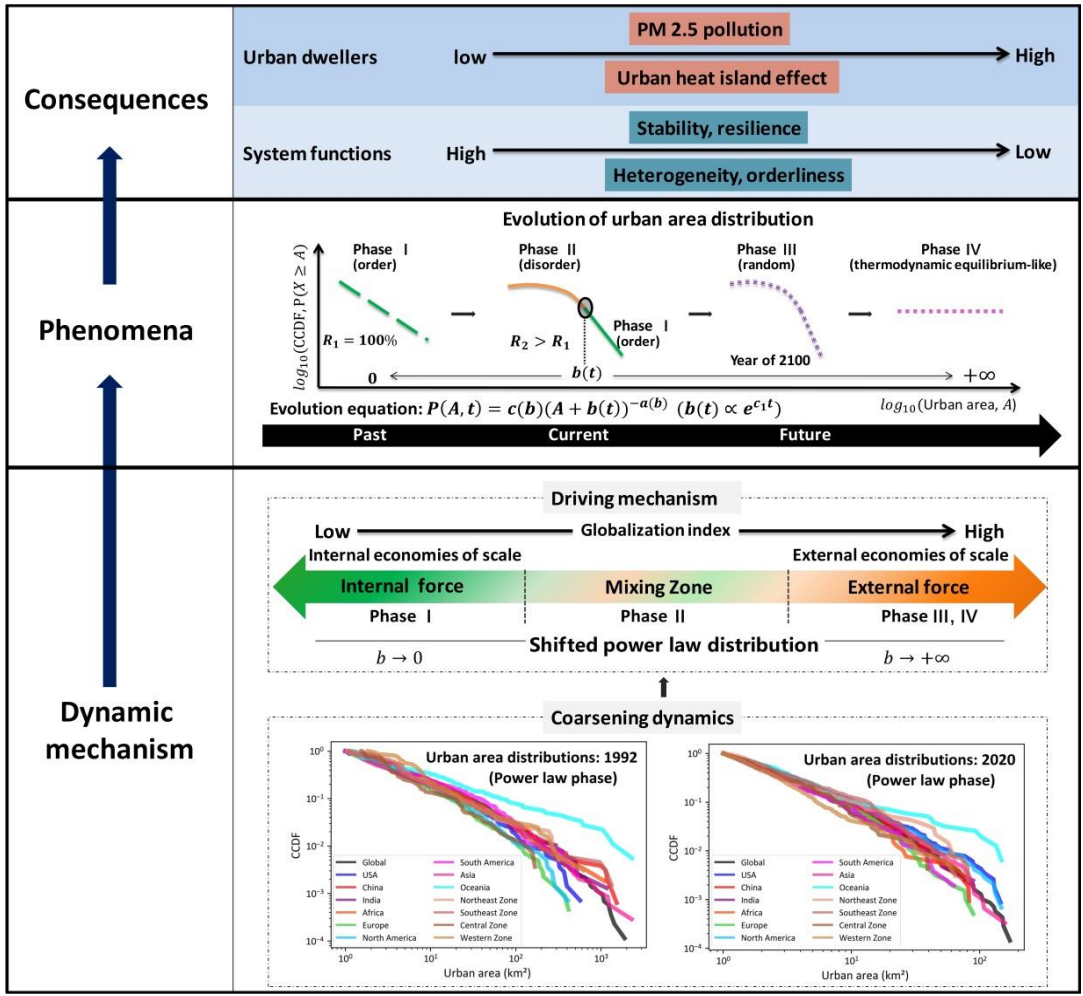

Fig. 1. Evolutionary process of global urban land expansion: statistical characteristics (phenomena), underlying mechanisms (dynamic mechanism), and impacts on urban systems and their inhabitants (consequences). **Phenomena:** Urban land data show that the urban area distribution has shifted from the initial power law distribution (Phase I) to the current state where the stretched-exponential (Phase II) and power law phase coexist, and will approach the exponential (Phase III) and uniform distribution (Phase IV) in the future. **Dynamic mechanism:** The shift in the urban area distribution can be attributed to the changing role of internal and external economies of scale associated with globalization. The driving mechanism is as follows: when internal economies of scale (internal force) dominate the urban system, urban areas follow the power law distribution; while as the influence of external economies of scale (external force) intensifies, the stretched-exponential phase appears and expands in the urban area distribution causing the power law phase to shrink. When external economies of scale (external force) become dominant, the power law phase disappears and urban areas are likely to enter the regime of exponential and further uniform distribution. Coarsening dynamics may be the dynamical principle governing this driving mechanism, as reflected by the dynamic scaling phenomenon in the study regions. **Consequences:** The shift in the urban area distribution has led to a decrease in the stability and resilience of urban systems and an increase in the exposure of urban dwellers to extreme heat events and air pollution.

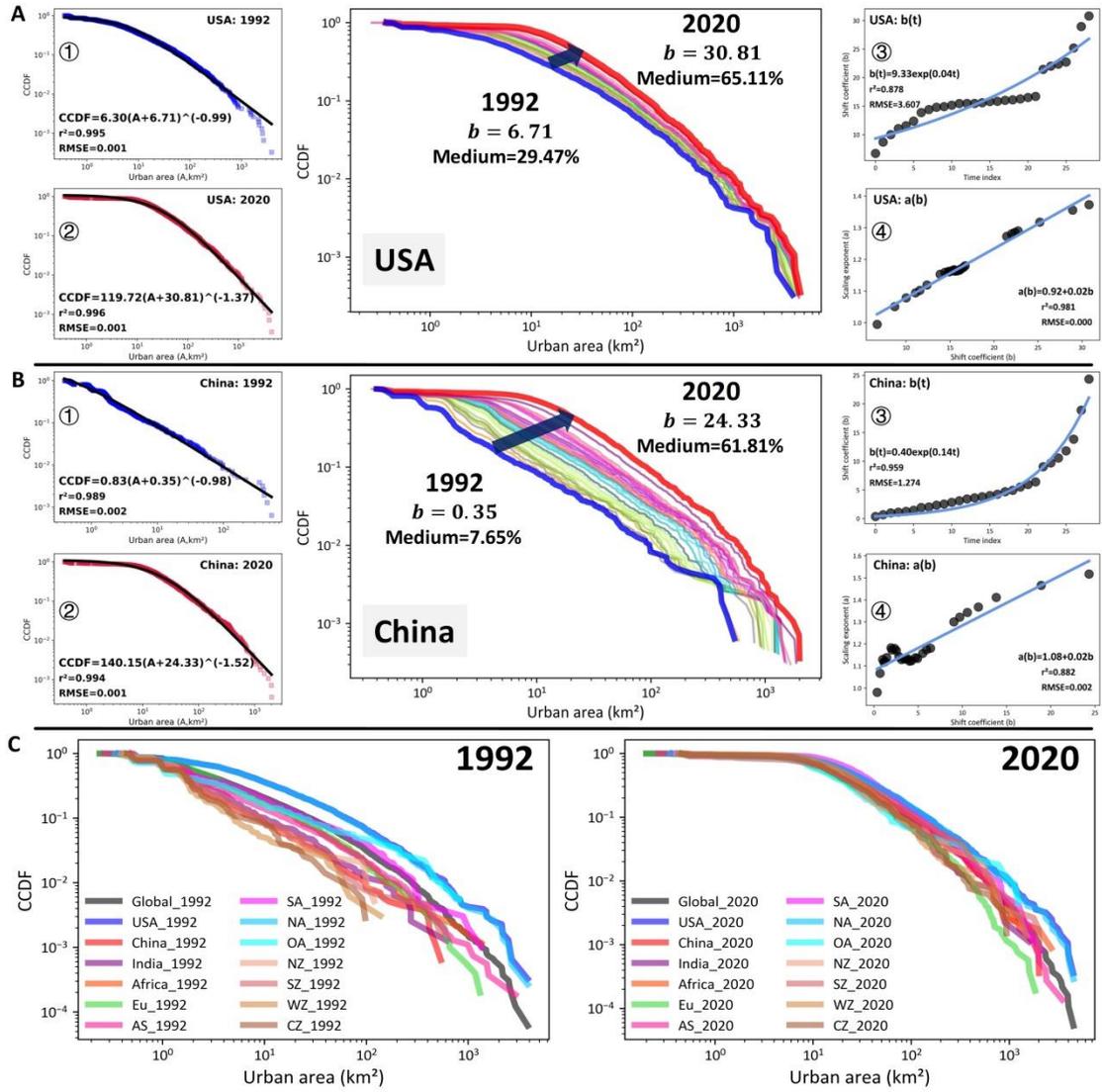

**Fig. 2. Converging trend of urban area distributions among different regions.** (**A**)-(**B**) Evolutionary characteristics of urban area distributions in USA and China, respectively. ①-②: Fitting results of the shifted power law distribution to urban areas in 1992 and 2020, respectively; ③: Exponential increase of the shifted coefficient ($b$) over time (here $t$ = Year-1992 is the time index); ④: Linear relationship between the scaling exponent ($a$) and the shifted coefficient ($b$); "Medium" denotes the proportion of medium-size cites ($10^1 \leq A < 10^2 \, \text{km}^2$) in the city configuration. (**C**) Cross-scale convergent evolution of urban area distributions.

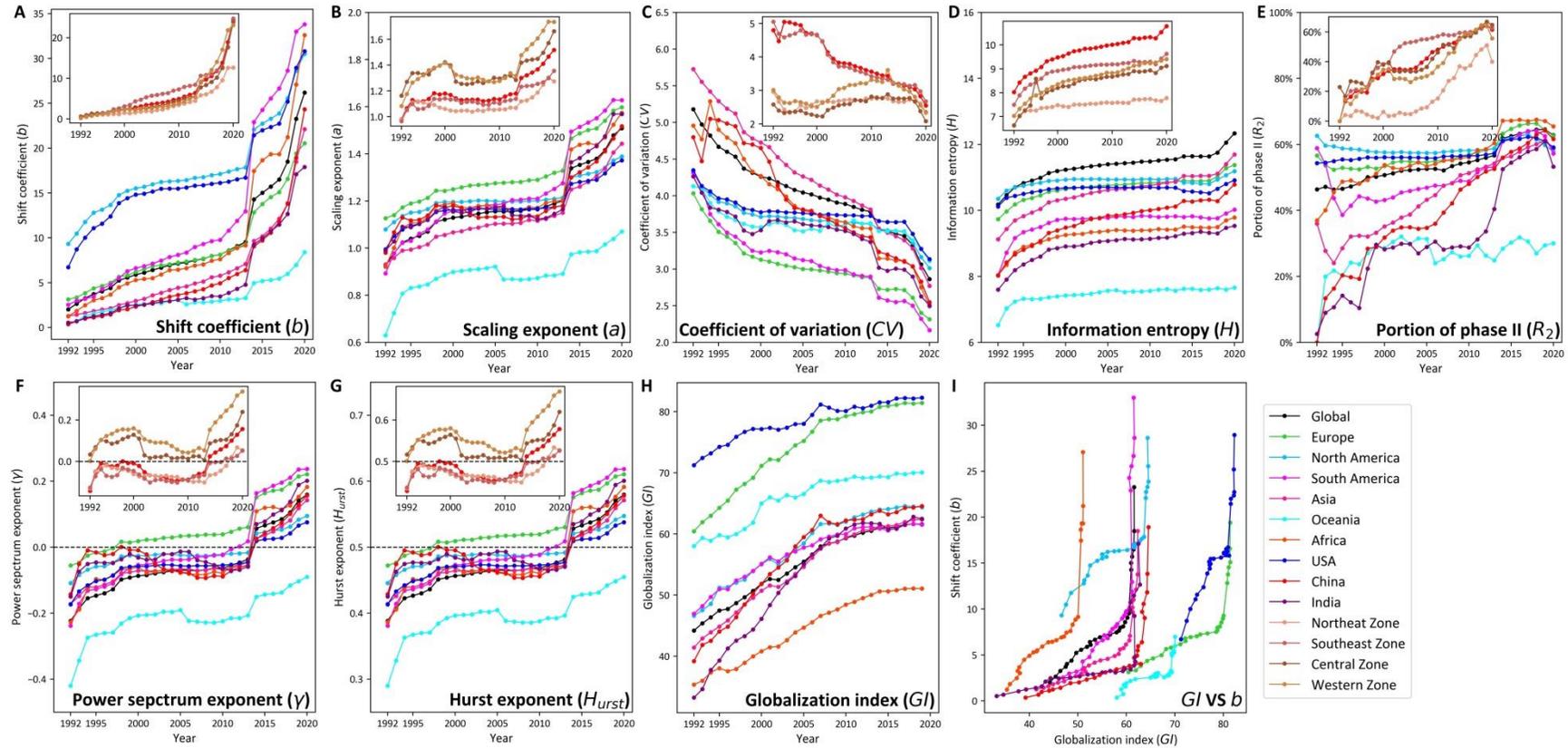

**Fig. 3. Temporal variations of indicators of urban system changes.** (**A**)-(**B**) Coefficients of urban area distributions; (**C**)-(**E**) State indicators of urban systems (Phase II: the stretched-exponential distribution); (**F**)-(**G**) Power spectrum variables; (**H**)-(**I**) Influence of external economies of scale. The enhanced influence from external economies of scale is represented by the increasing values of globalization index (GI) (**H**) and the positive correlations between GI values and the shifted coefficient (*b*) values in all study regions (**I**). The subplots show the results for China and its four economic zones.

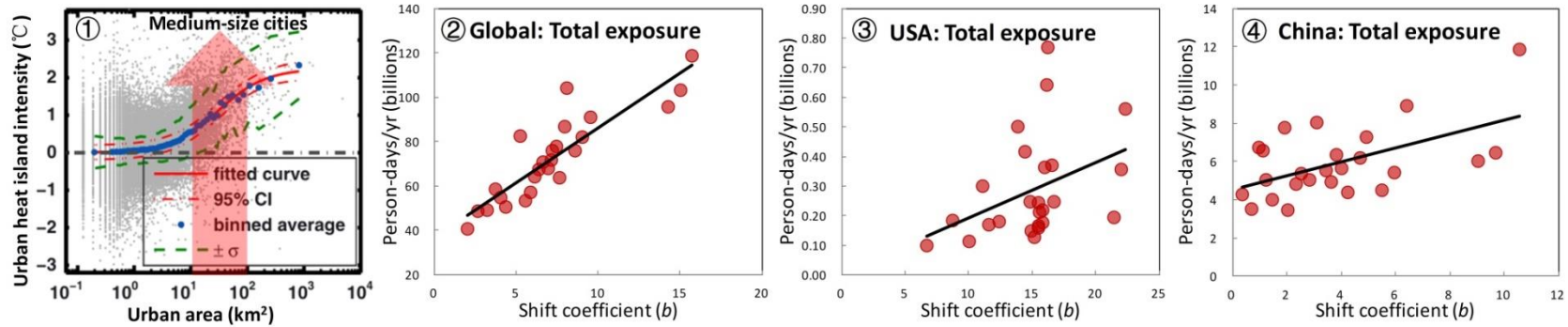
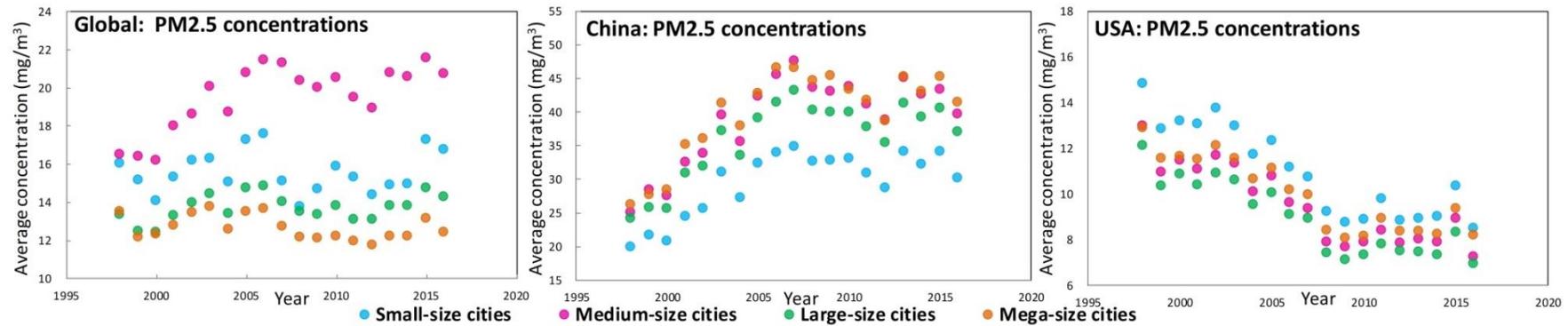

**Fig. 4. Impacts of urban system alterations on urban dwellers.** (**A**) Urban heat island effect (UHI): ①: Relationship between the UHI intensity and urban land area (Zhou et al., 2013); ②-④: Positive correlation between the total exposure of urban populations to extreme heat and the shift coefficient (*b*). (**B**) Fine particulate matter (PM2.5) pollution: average PM2.5 concentrations for each city type.

**Table 1. Pattern shift in the city configuration experienced by all regions.** From small-size cities dominated to medium-size ones dominated. EU and SE are abbreviations for Europe and Southeast economic zone of China, respectively.

| Type | Global | | EU | | USA | | China | | India | | China_SZ | |
|---|---|---|---|---|---|---|---|---|---|---|---|---|
| | 1992 | 2020 | 1992 | 2020 | 1992 | 2020 | 1992 | 2020 | 1992 | 2020 | 1992 | 2020 |
| Small | 79.27% | 29.80% | 81.65% | 37.28% | 63.87% | 20.79% | 91.59% | 29.30% | 89.95% | 30.54% | 91.49% | 33.83% |
| Medium | 17.60% | 60.24% | 16.45% | 55.64% | 29.47% | 65.11% | 7.65% | 61.81% | 8.91% | 62.53% | 7.90% | 55.57% |
| Large | 2.98% | 9.56% | 1.86% | 6.99% | 6.24% | 13.11% | 0.76% | 8.57% | 1.15% | 6.82% | 0.61% | 10.16% |
| Mega | 0.15% | 0.41% | 0.04% | 0.09% | 0.42% | 1.00% | 0.00% | 0.32% | 0.00% | 0.11% | 0.00% | 0.44% |

**Table 2. Value variations of distribution function coefficients, state indicators and power spectrum variables between 1992 and 2020.** Distribution function coefficients: shift coefficient ($b$) and scaling exponent ($a$); State indicators: coefficient of variation (CV), information entropy (H) and portion of the stretched-exponential phase ($R_2$); Power spectrum variables: spectrum exponent ($\gamma$) and Hurst exponent ($H_{urst}$). EU: Europe; NA: North America; OA: Oceania; AF: Africa; AS: Asia; SA: South America; NZ, SZ, CZ and WZ are the Northeast, Southeast, Central and Western economic zone of China, respectively.

| Regions | Distribution function coefficients | | State indicators | | | Power spectrum variables | |
|---|---|---|---|---|---|---|---|
| | $b$ | $a$ | CV | H | $R_2$ | $\gamma$ | $H_{urst}$ |
| Global | 24.18 | 0.58 | -2.32 | 2.16 | 16.44% | 0.38 | 0.19 |
| EU | 17.42 | 0.47 | -1.72 | 1.65 | 5.97% | 0.28 | 0.14 |
| NA | 21.19 | 0.31 | -1.33 | 0.84 | -4.85% | 0.20 | 0.10 |
| OA | 8.04 | 0.44 | -1.02 | 1.14 | 30.04% | 0.33 | 0.16 |
| AF | 31.38 | 0.65 | -2.46 | 1.75 | 28.53% | 0.41 | 0.20 |
| AS | 20.85 | 0.51 | -2.96 | 2.57 | 22.82% | 0.31 | 0.16 |
| SA | 31.27 | 0.74 | -2.15 | 1.98 | -1.64% | 0.47 | 0.24 |
| USA | 24.1 | 0.38 | -1.21 | 0.80 | 4.78% | 0.25 | 0.12 |
| India | 17.36 | 0.58 | -1.76 | 1.93 | 50.70% | 0.35 | 0.18 |
| China | 23.98 | 0.54 | -2.25 | 2.76 | 61.60% | 0.30 | 0.15 |
| NZ | 12.63 | 0.3 | -0.63 | 1.14 | 39.91% | 0.18 | 0.09 |
| SZ | 24.69 | 0.39 | -2.38 | 2.14 | 62.72% | 0.18 | 0.09 |
| CZ | 23.88 | 0.5 | -0.50 | 2.49 | 41.75% | 0.21 | 0.10 |
| WZ | 23.12 | 0.65 | -0.66 | 2.38 | 55.63% | 0.34 | 0.17 |

**Table 3. Future trajectories of urban land expansion in different regions.** Shift coefficient, $b$; portion of power law phase, $R_1$.

| Future | | Global | EU | NA | SA | AS | OA | AF | USA | China | India |
|---|---|---|---|---|---|---|---|---|---|---|---|
| 2050 | $b$ | 2.89E+02 | 1.34E+02 | 7.17E+01 | 7.00E+02 | 5.40E+02 | 5.28E+01 | 8.15E+02 | 8.30E+01 | 1.47E+03 | 9.32E+02 |
| | $R_1$ | 2.96% | 4.80% | 20.75% | 1.00% | 1.17% | 12.11% | 0.35% | 17.36% | 0.14% | 0.11% |
| 2100 | $b$ | 1.96E+04 | 3.56E+03 | 3.69E+02 | 1.03E+05 | 1.41E+05 | 1.50E+03 | 2.08E+05 | 5.46E+02 | 1.75E+06 | 6.90E+05 |
| | $R_1$ | 0.00% | 0.00% | 3.50% | 0.00% | 0.00% | 0.00% | 0.00% | 2.14% | 0.00% | 0.00% |